\documentclass[11pt]{article} \textwidth 170mm \textheight 225mm
\usepackage{amsfonts}
\usepackage{amssymb}
\usepackage{dcolumn}
\usepackage{amsthm}
\usepackage{amsmath}
\usepackage{multicol}
\usepackage{newlfont}
\usepackage{newlfont}
\usepackage{multicol}
\usepackage{bm}
\usepackage{txfonts}
\usepackage{graphics}
\usepackage{graphicx}
\usepackage{color}
\usepackage{indentfirst}
\begin{document}

\title{The temperature-dependent elastic properties of B2-MgRE intermetallic compounds  from first principles \footnote{The work is supported by the National Natural Science Foundation
of China (11074313) and and Project No.CDJXS11102211 supported by the Fundamental Research Funds for the Central Universities of China. }}
\author{Rui Wang\footnote{Tel: +8613527528737; E-mail: rcwang@cqu.edu.cn.}, Shaofeng Wang, and Yin Yao\\
{\small  { Institute for Structure and Function and
department of physics, Chongqing University, }}\\ {\small {Chongqing 400044, People's Republic of China. }} }

\date{}

\maketitle
\begin{abstract}
\baselineskip 18pt
\noindent Using the density functional theory (DFT) formulated within the framework of the plane-wave basis projector augmented wave (PAW) method,  the temperature-dependent elastic properties of MgRE (RE=Y, Dy, Pr, Sc, Tb) intermetallics  with B2-type structure are presented from  first-principles. Our calculations are based on the fact that the elastic moduli as a function of temperature mainly results from thermal expansion. The comparison between the predicted results and the available experimental data for a benchmark material NiAl provides good agreements. At $T=0K$, our calculated values of lattice parameter and elastic moduli for MgRE intermetallics show excellent agreement with previous theoretical results and experimental data. While temperature increases, we find that the elastic constants decrease and approach linearity at higher temperature and zero slope around zero temperature.
\end{abstract}

\vskip 0.1in{\small   } \vskip 0.2in

\vskip 0.1in

 \vskip 0.3in

\baselineskip 20pt


\section{Introduction}

Since magnesium (Mg) alloy often shows high strength to weight ratio \cite{Kainer}, it is one of the potential materials for technological applications. Especially, Mg alloys have been attractive for the applications of aeronautical and automotive industry. However, the low strength and creep resistance at high temperature is a very serious problem in application of magnesium alloys. Therefore, it is quite necessary to improve the strength and creeping resistance of Mg-based alloys at elevated temperatures. Recently, it has been reported that some rare-earth-Mg intermetallic compounds MgRE (where RE indicates a rare-earth element) with B2-type structure have good creep and high temperature strength \cite{Mordike1,Mordike2,Lorimer}. MgRE intermetallics are extremely attractive structural materials for applications in automobile parts and aerospace industries. So various properties such as magnetic properties, linear and nonlinear elasticity, stacking fault, and thermal properties for the B2-MgRE intermetallics have been intensely investigated \cite{Luca,Deldem,Wang2010,Wang2011p,Wu,Tao,Wang2011,Guo,Cacciamani,Wu1}. In our previous study, the temperature dependence of various quantities such as the thermal expansions, bulk modulus, and the heat capacity are obtained by using the density functional theory (DFT) and density functional perturbation theory (DFPT) \cite{Wang2011}.

In general, elastic properties of a solid are very important because they are closely associated with various fundamental solid-state properties such as interatomic potentials, equation of state, and phonon spectra. The temperature dependence of the elastic constants of a material is important for predicting and understanding the mechanical strength, stability, and phase transitions of a material \cite{Gulsern}. Though modern electronic-structure methods based on DFT can treat the zero-temperature pressure dependence elastic moduli from first principles accurately \cite{Karki}, treating the corresponding temperature dependence of elastic constants is still a formidable challenge \cite{Orlikowski}. However, the temperature-dependent elastic properties have also been investigated by using the first-principles DFT methods in several groups \cite{Gulsern,Orlikowski,Kadas,Sha,Wang,Shang}. Especially, Wang et. al. \cite{Wang,Shang} demonstrate that the temperature dependence of elastic moduli mainly results from volume change as a function of temperature, and their applications to many materials show excellent agreement between the calculated values and experimental data. In this paper, we employ the first-principles method \cite{Wang,Shang} to study the temperature-dependent elastic constants for MgRE (RE=Y, Dy, Pr, Sc, Tb) intermetallics with B2-type structure. To judge that our
computational accuracy is reasonable, we have also computed the temperature dependence of elastic moduli for a benchmark material NiAl, which has the same B2-type structure and accompanied by available experimental data.

\section{Theoretical methods}

The Helmholtz free energy of intermetallics at a constant volume $V$ and $T$ has three major contributions \cite{Moriarty, Wang1}
\begin{equation}\label{Ftotal}
F(V, T)=E_{\mathrm{static}}(V)+F_{\mathrm{el}}(V,T)+F_{\mathrm{ph}}(V, T),
\end{equation}
where $E_{\mathrm{static}}$ is the zero-temperature energy of a static lattice, $F_{\mathrm{el}}$ is the thermal electronic contribution to free energy from finite temperature, and $F_{\mathrm{ph}}$ represents the phonon free energy arising from the lattice vibrations. We obtain both $E_{\mathrm{static}}$ and $F_{\mathrm{el}}$ from first-principles DFT calculations directly, assuming that the eigenvalues for given lattice and ion positions are temperature independent and only the occupation numbers change with temperature through the Fermi-Dirac distribution \cite{Sha,Wasserman}. The phonon free energy within the framework of quasiharmonic approach (QHA) is obtained from
\begin{equation}\label{Fph}
F_{\mathrm{vib}}(V, T)=\sum_{\kappa}\Bigg[\frac{1}{2}{\hbar\omega_{\kappa}}+{k_{B}T}\ln\bigg(1-e^{-{\hbar\omega_{\kappa}}/{k_{B}T}}\bigg)\Bigg],
\end{equation}
where $\omega_{\kappa}$ represents an individual phonon frequency.

The isothermal elastic constants are derived from the second-order strain derivative of the Helmhotz free energy  \cite{Brugger}
\begin{equation}\label{Cijkl}
C_{ijkl}^{T}=\frac{1}{V}\frac{{\partial}^{2}F}{\partial \varepsilon_{ij}
\partial \varepsilon_{kl}}\bigg|_{T,\varepsilon '},
\end{equation}
where $\varepsilon '$ indicates that all other stains are held fixed. The elastic behaviour of B2-MgRE intermetallics is completely described by three independent constants $C_{11}^{T}$, $C_{12}^{T}$, and $C_{44}^{T}$ (in Voigt notation). The bulk modulus is defined by a linear combination of the elastic constants
\begin{equation}\label{Bulk}
B_{T}=(C_{11}^{T}+2C_{12}^{T})/3,
\end{equation}
and determined from the Vinet equation of state \cite{Vinet} corresponding to a pure volume deformation of the lattice. We apply a tetragonal strain to calculate $(C_{11}^{T}-C_{12}^{T})$
\begin{equation}\label{strain1}
\varepsilon(\delta)=\left( \begin{array}{ccc}
                      \delta & 0 & 0 \\
                      0 & \delta & 0 \\
                      0 & 0 & (1+\delta)^{-2}-1
                    \end{array}
\right),
\end{equation}
and the corresponding free energy  is
\begin{equation}\label{Fs1}
F(V,\delta)=F(V,0)+3(C_{11}^{T}-C_{12}^{T}) V \delta^{2}+O(\delta^{3}),
\end{equation}
where $F(V,0)$ is the free energy of the unstrained state.
$C_{44}^{T}$ is obtained from a volume-conserving monoclinic stain,
\begin{equation}\label{strain2}
\varepsilon(\delta)=\left( \begin{array}{ccc}
                      0 & \delta & 0 \\
                      \delta & 0 & 0 \\
                      0 & 0 & \delta^{-2}/(1-\delta^{-2})
                    \end{array}
\right),
\end{equation}
which leads to the free energy change
\begin{equation}\label{Fs2}
F(V,\delta)=F(V,0)+2C_{44}^{T} V \delta^{2}+O(\delta^{4}),
\end{equation}
We calculate the free energy for different strains, and then the elastic constants are calculated from the fitted quadratic coefficients. To compare with experiment, the isothermal elastic moduli must also be transformed to the adiabatic elastic moduli by the following relation \cite{Davies}
\begin{equation}\label{TtoS}
C_{ij}^{S}=C_{ij}^{T}+\frac{TV}{C_{V}}\sum_{kk'}\alpha_{k}\alpha_{k'}C_{ik}^{T}C_{jk'}^{T}
\end{equation}
where $C_{V}$ is the specific heat at constant volume and $\alpha_{k}$ is the linear thermal expansion tensor. For B2-MgRE intermetallics with cubic crystal, Eq. (\ref{TtoS}) simplifies to \cite{Orlikowski}
\begin{eqnarray} \label{CTS}
&C_{44}^{S}=C_{44}^{T},\nonumber \\
&C_{11}^{S}-C_{11}^{T}=C_{12}^{S}-C_{12}^{T}=\frac{TV}{C_{V}}\alpha^{2}B_{T}^{2},
\end{eqnarray}
with $\alpha$ is the thermal expansion coefficient. From Eqs. (\ref{Bulk}) and (\ref{CTS}), it follows that the adiabatic bulk modulus is just
\begin{equation} \label{BTS}
B_{S}=B_{T}+\frac{TV}{C_{V}}\alpha^{2}B_{T}^{2}
\end{equation}

We use a quasistatic approach, which was recently developed by Wang et al.\cite{Wang,Shang}, to calculate the temperature-dependent elastic constants $C_{ij}(T)$. In the quasistatic method, it is assumed that the temperature dependence of elastic constants is solely caused by thermal expansion. The applications of quasistatic approach to elastic constants of many materials from 0K up to their melting points show excellent agreement between the computed values and experimental data \cite{Wang,Shang}. In this paper, the temperature-dependent elastic constants are obtained in the following three steps \cite{Wang}. The first step is calculating the thermal expansion and the equilibrium volume  $V(T)$ at $T$  by using the first-principles quasiharmonic approximation, while isothermal bulk modulus $B_{T}$ as a function of temperature $T$ is also obtained from fitting to the Vinet equation of state \cite{Vinet}. The detailed procedure of calculating the thermal expansion coefficient, isothermal bulk modulus $B_{T}$ were shown in our recent work \cite{Wang2011}. In the second step, we predict the static elastic constants as a function of volume $C_{ij}^{T}(V)$
by using the energy-strain relation based on Eqs.(\ref{Bulk}-\ref{Fs2}). In the third step, the calculated elastic constants from the second step at the volume $V(T)$ are approximated as those at finite temperatures, i.e., $C_{ij}^{T}(T)=C_{ij}^{T}[T(V)]$. Then, the adiabatic elastic constants $C_{ij}^{S}$ can be obtained from Eqs. (\ref{CTS}) and (\ref{BTS}).

The computational approach is based on the density-functional theory (DFT) and density-functional perturbation theory (DFPT) as implemented in the VASP package \cite{Kresse1, Kresse2,Kresse3,Kresse5}. We use the Perdew-Burke-Ernzerhof (PBE) \cite{Perdew1,Perdew2} generalized gradient approximation (GGA) for the exchange-correlation functional. The plane-wave basis projector augmented wave (PAW) method \cite{Blochl, Kresse4} is used. Since high accuracy is needed to evaluate the elastic constants, the convergence of strain energies with respect to the Brillouin zone integration was carefully checked by repeating the calculations for $21\times21\times21$ and $25\times25\times25$  Monkhorst-Pack \cite{Monkhorst} meshes, and we found that fluctuation both for $C_{44}$ and $\mu$ is lower than 0.5GPa. Hence, we used $21\times21\times21$ in the full Brillouin zone giving 726 irreducible k-points. In addition, we used a high plane-wave energy cutoff of 600eV which is sufficient to calculate the elastic moduli accurately.

To calculate the phonon frequency,  we have carried out DFPT calculations on this $3\times3\times3$ supercell with 54 atoms by using PBE-GGA exchange-correlations effects and $7\times7\times7$ k-point grid meshes for Brillouin zone integrations. The phonon free energy were obtained by using PHONOPY \cite{Togo2010,Togo2008,Togop} package which can support VASP interface to calculate force constants matrix directly. To consider the thermal expansion, we calculate the vibrational frequencies of all calculated intermetallics at 13 volumes.  The thermal expansion coefficients $\alpha$ and the thermal bulk modulus $B_{T}$ are obtained by minimizing Helmholtz free energy Eq. (\ref{Ftotal}) with respect to $V$ from fitting the integral form of the Vinet equation of state (EOS) \cite{Vinet}.

\section{Results and discussion}

The calculated equilibrium lattice constants $a$, elastic constants $C_{ij}$, and bulk modulus $B$ at $T=0K$ for  MgRE (RE=Y, Dy, Pr, Sc, Tb) in comparison with the previous calculated results \cite{Wu,Tao} are listed in Table \ref{table}. Where available, the  experimental data \cite{Villars} for lattice constants are also listed for comparison. The calculated lattice constants are in excellent agreement with  experimental data within 0.5\%. While the lattice constants in present calculations shows no difference from the results obtained by Tao et. al.\cite{Tao}, since we employ the PBE-GGA exchange-correlation functional compared to PW91-GGA in their calculation. No temperature-dependent elastic constants have been reported theoretically and experimentally for all MgRE intermetallics. At $T=0K$, our calculated results for elastic constants agree well with the previous DFT data \cite{Wu,Tao} and it is noticeable that, the requirement of mechanical stability for all calculated MgRE, namely,  $C_{11}-C_{12}>0$, $C_{11}>0$, and $C_{44}>0$ \cite{Nye}, is satisfied.

\begin{figure}
\scalebox{0.6}[0.6]{\includegraphics{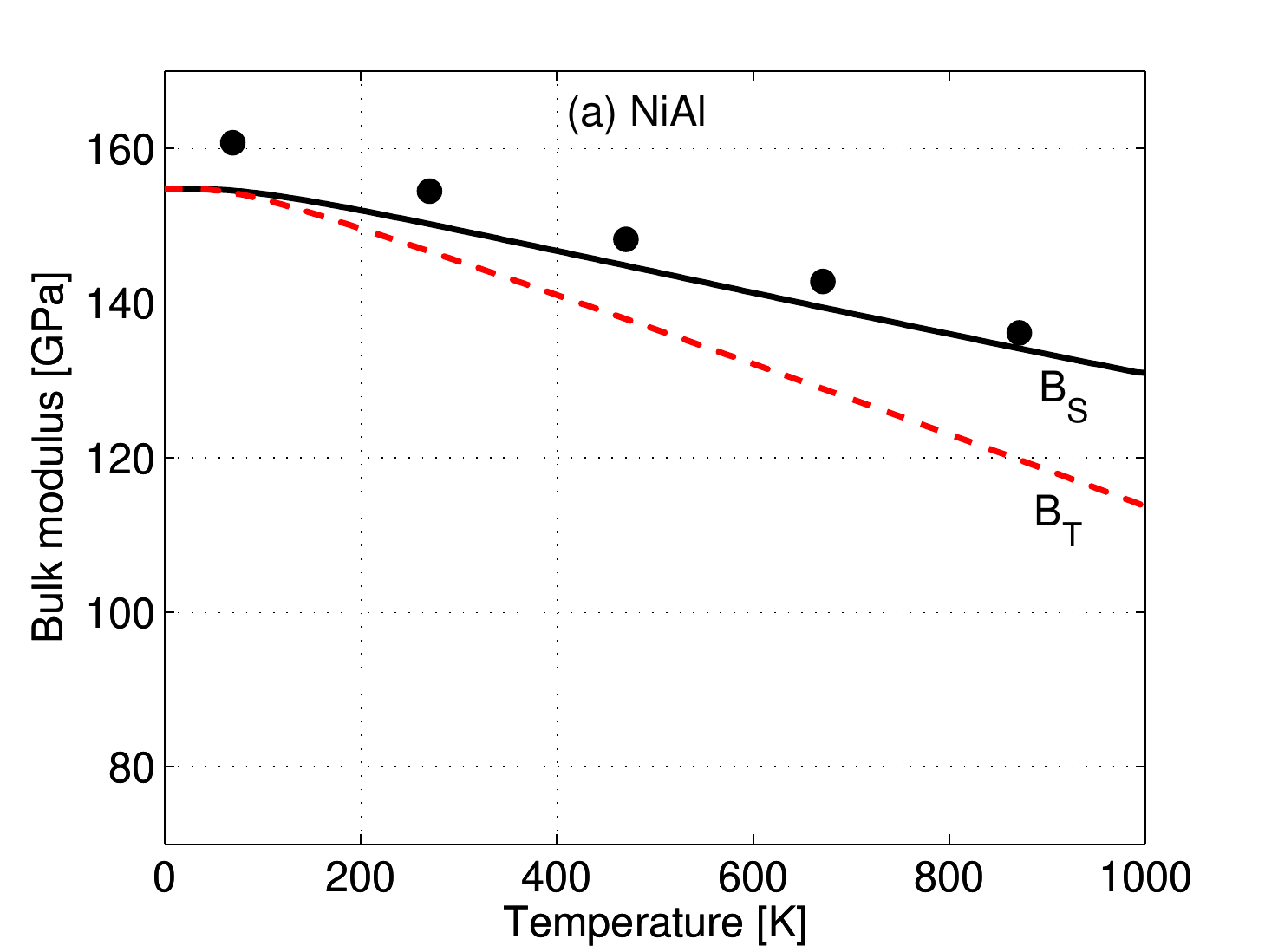}}
\scalebox{0.6}[0.6]{\includegraphics{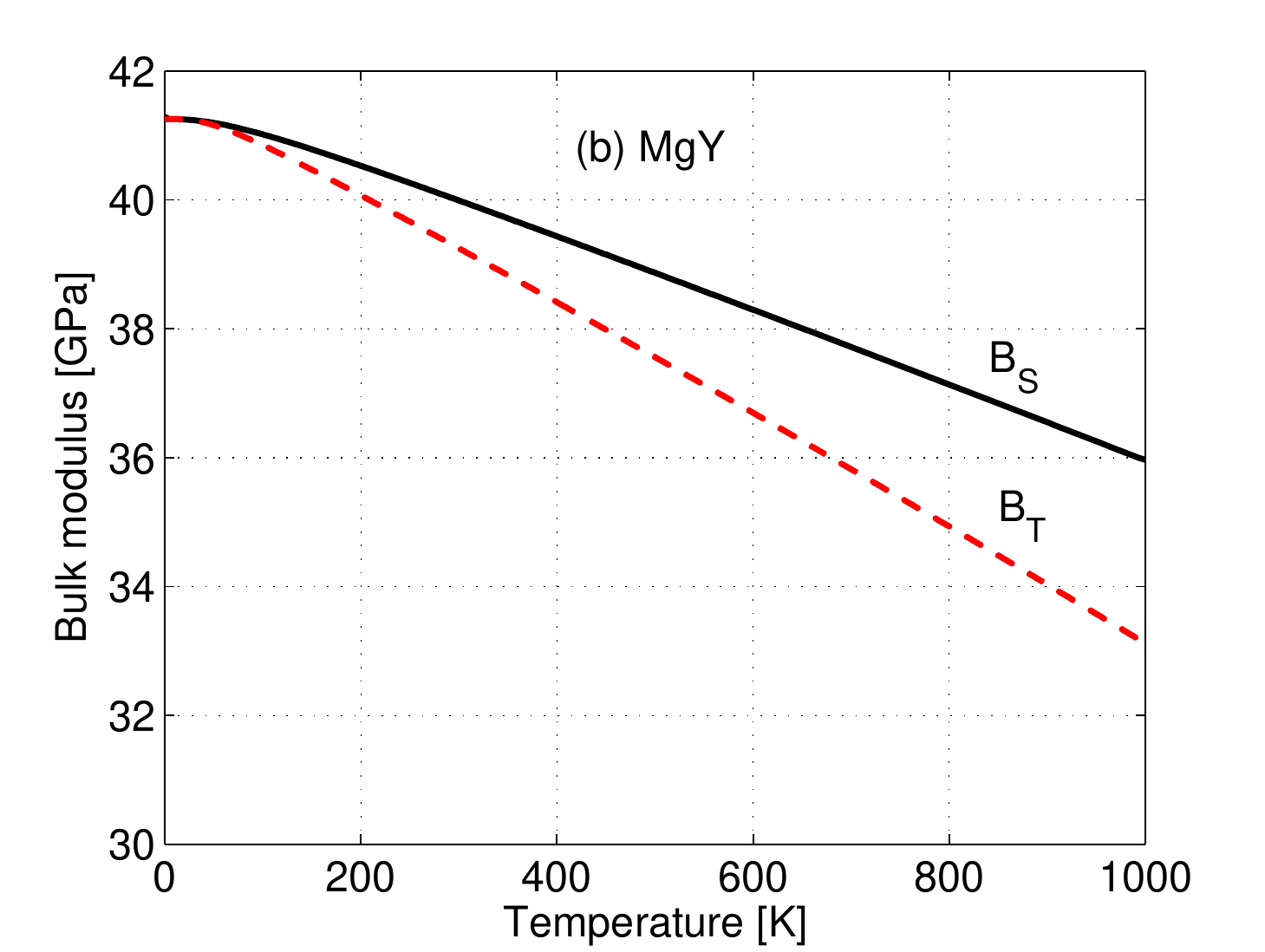}}
\scalebox{0.6}[0.6]{\includegraphics{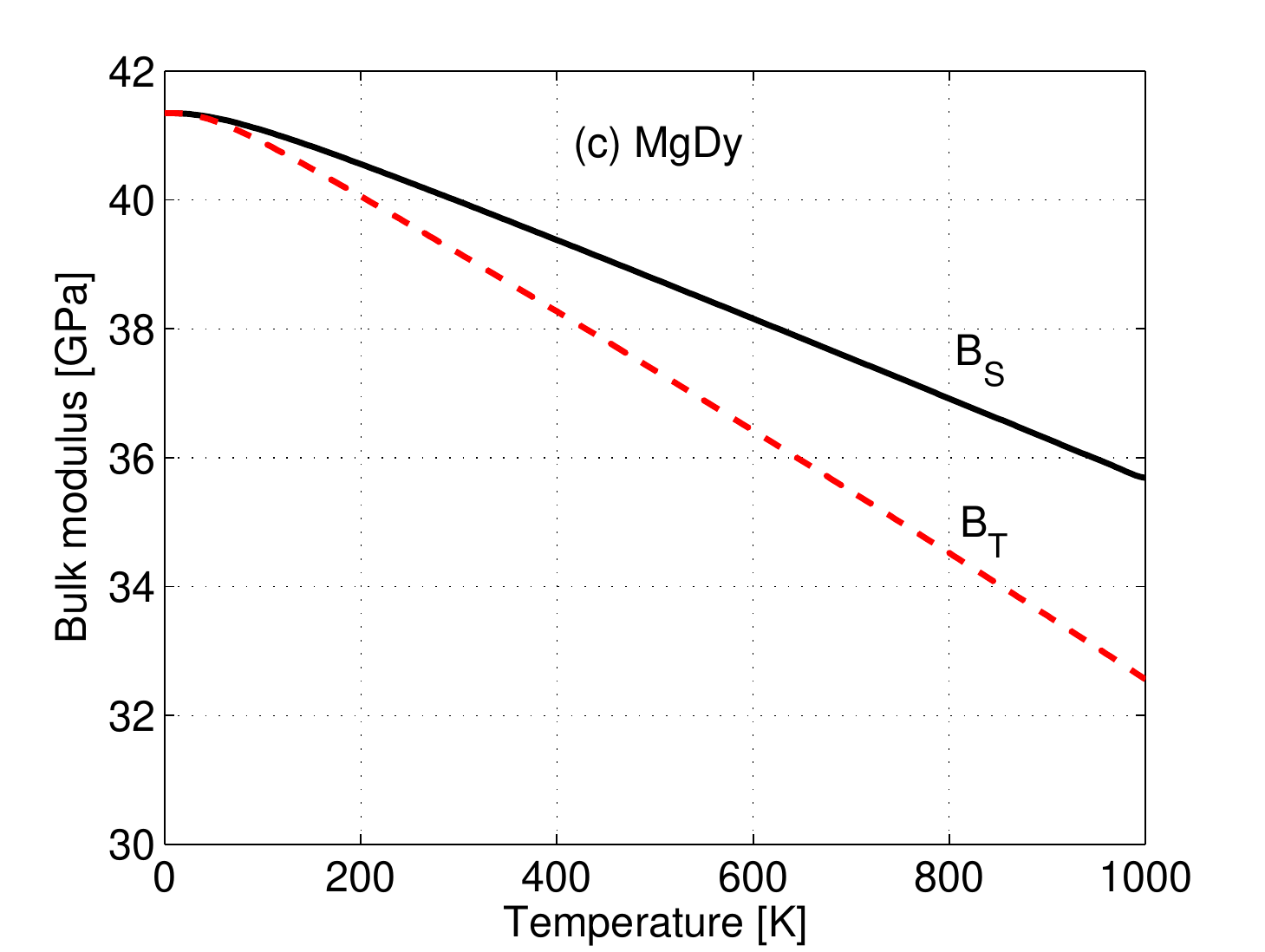}}
\scalebox{0.6}[0.6]{\includegraphics{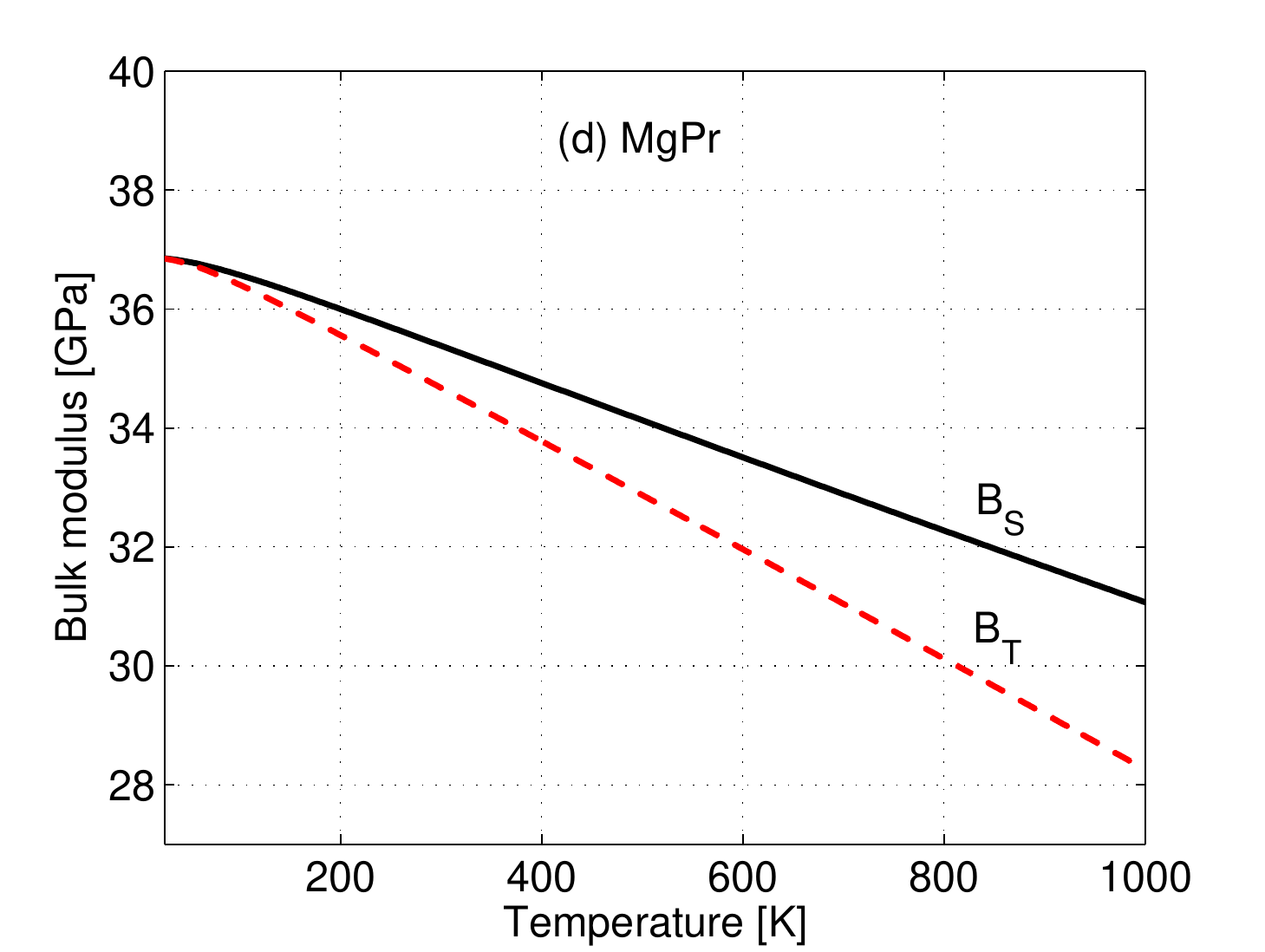}}
\scalebox{0.6}[0.6]{\includegraphics{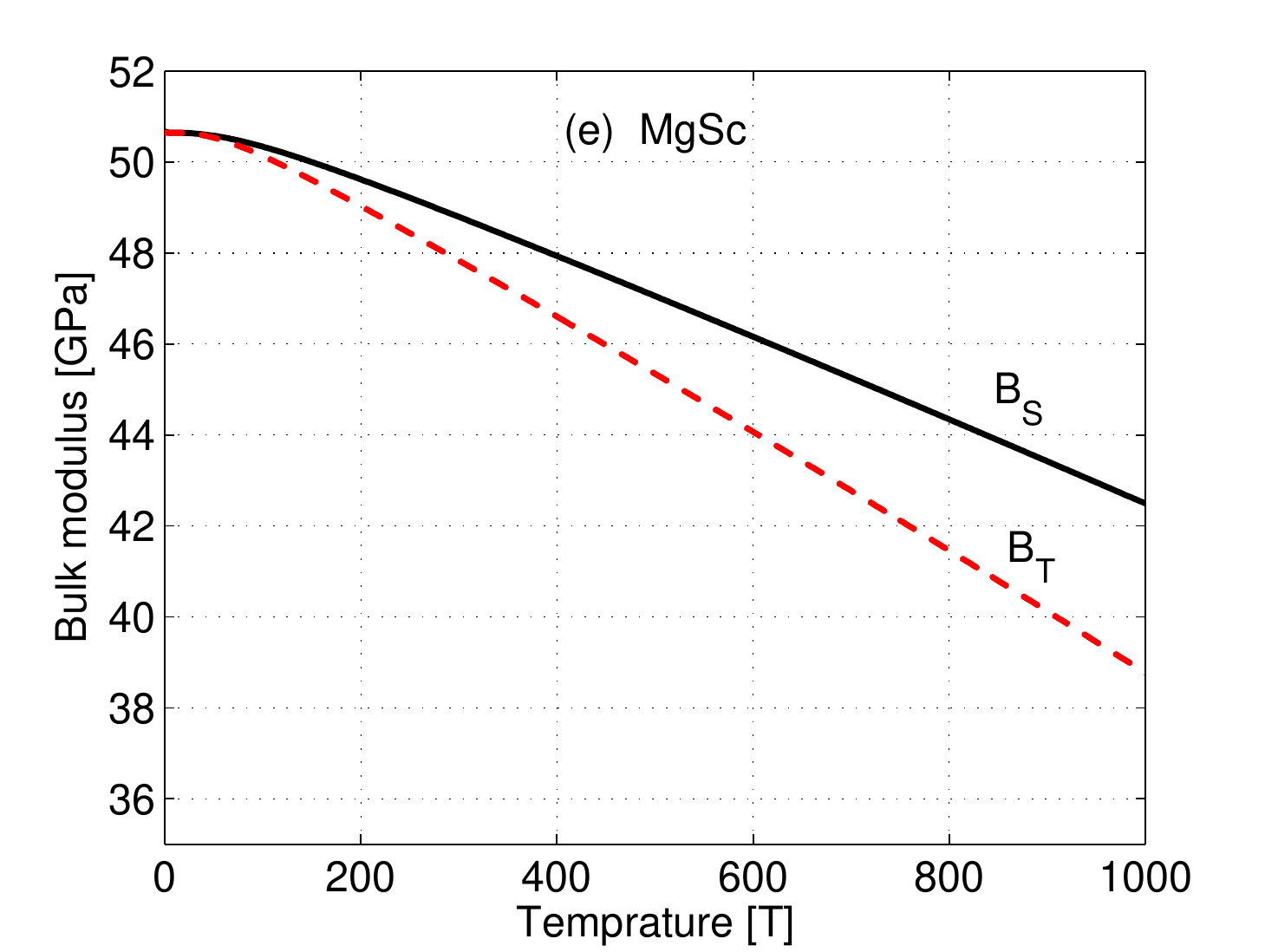}}
\scalebox{0.6}[0.6]{\includegraphics{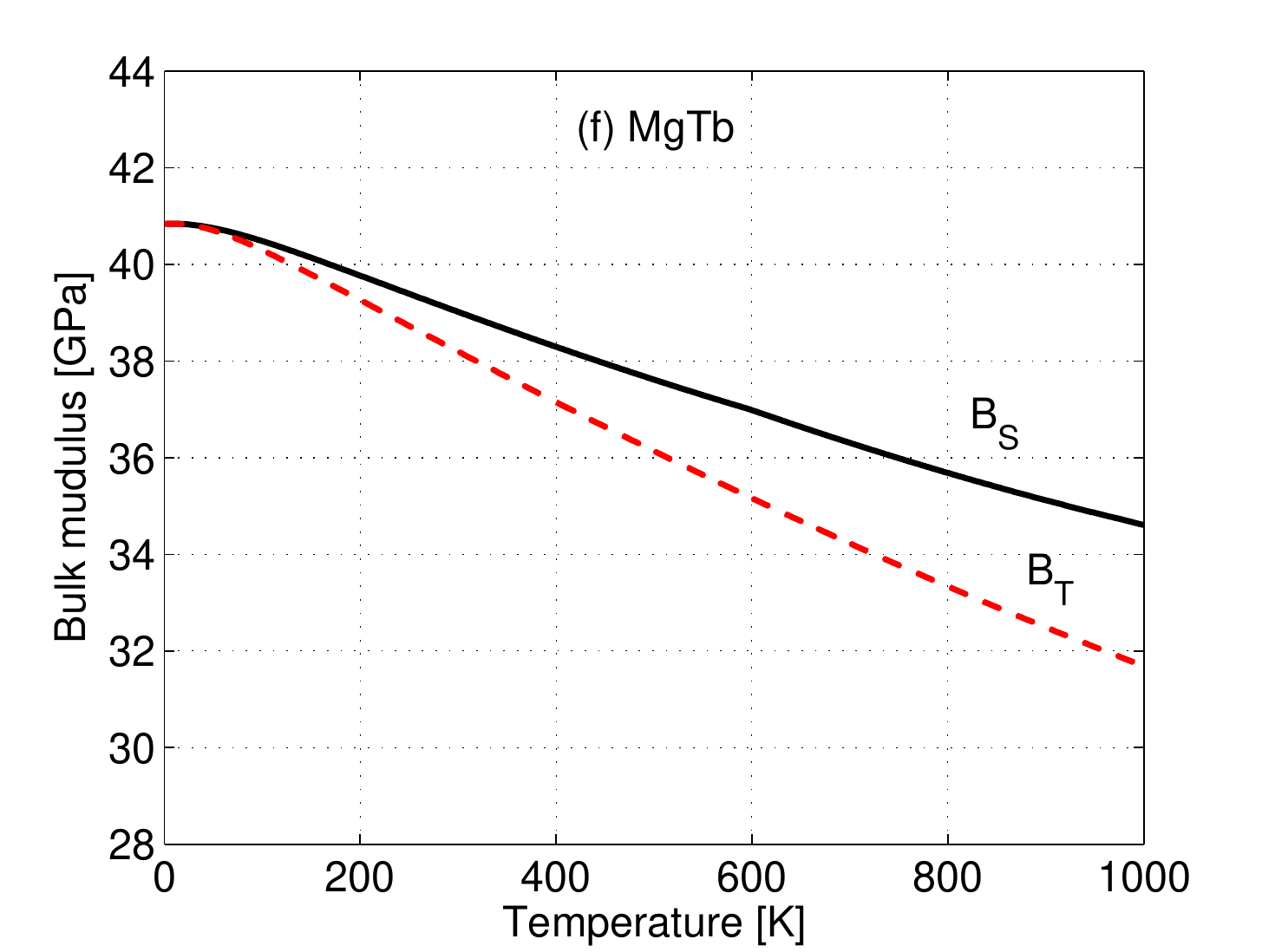}}
\caption{ \footnotesize {(Color on line) The temperature-dependent adiabatic bulk modulus $B_{S}$ (solid line) as well as the calculated isothermal bulk modulus $B_{T}$ (dashed line) for (a) NiAl, (b) MgY, (c) MgDy, (d) MgPr, (e) MgSc, and (f) MgTb. For NiAl, the full symbols denote the corresponding experimental data form Rusovi\'{c} and Warlimont \cite{Rusovic}}. }
\label{bulk}
\end{figure}

The calculated isothermal and adiabatic bulk modulus for NiAl and MgRE (RE=Y, Dy, Pr, Sc, Tb) as a function of temperature are shown in Figure \ref{bulk}. The all isothermal bulk modulus $B_{T}$ are obtained by minimizing Helmholtz free energy Eq. (\ref{Ftotal}) with respect to $V$ from fitting the integral form of the Vinet equation of state (EOS) \cite{Vinet}. Figure \ref{bulk}(a) shows our findings for benchmark materials NiAl, which has the same B2-type structure as MgRE intermetallics, accompanied by available experimental data form Rusovi\'{c} and Warlimont \cite{Rusovic}. The calculated values of adiabatic bulk modulus $B_{S}$ for NiAl are in good agreement with experiments. Our results also show no discrepancy with previous DFT calculated values obtained  from using the quasistatic approach by Wang et. al. \cite{Wang}. The overall observation is that both the $B_{T}$ and $B_{S}$ decrease with increasing temperature and approach linearity at higher temperature and zero slope around zero temperature. It it worth to note that the corrections ${TV}\alpha^{2}B_{T}^{2}/{C_{V}}$ between adiabatic elastic moduli and isothermal ones increase with increasing temperature. Among the five intermetallics [shown in Figure \ref{bulk}(b)-(f)], through the temperature range, MgSc [Figure \ref{bulk}(e)] and MgPr [Figure \ref{bulk}(d)] have the highest and the lowest bulk moduli, respectively, i.e., MgSc is the most incompressible and MgPr is the most compressible.

\begin{figure}
\scalebox{0.6}[0.6]{\includegraphics{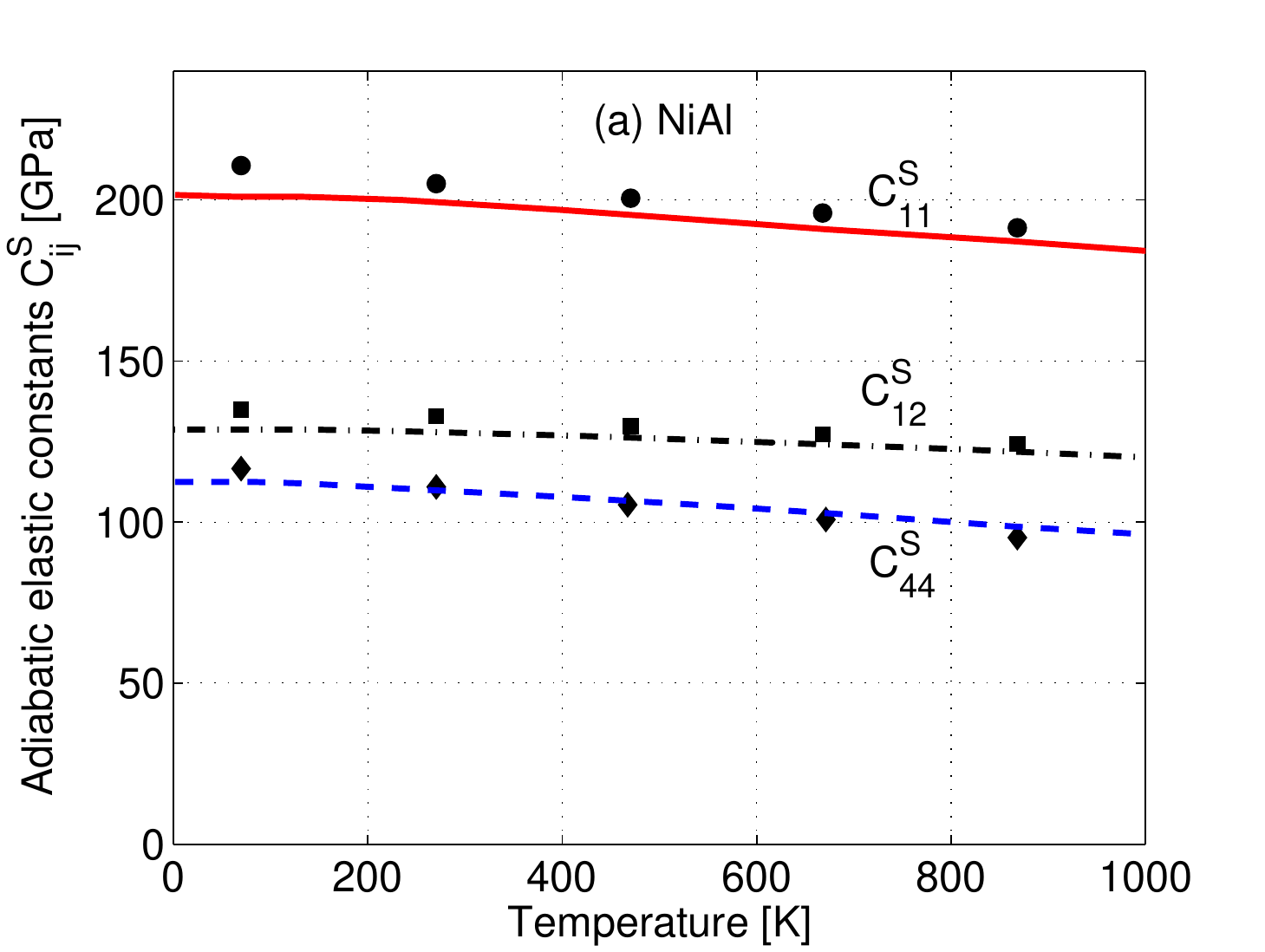}}
\scalebox{0.6}[0.6]{\includegraphics{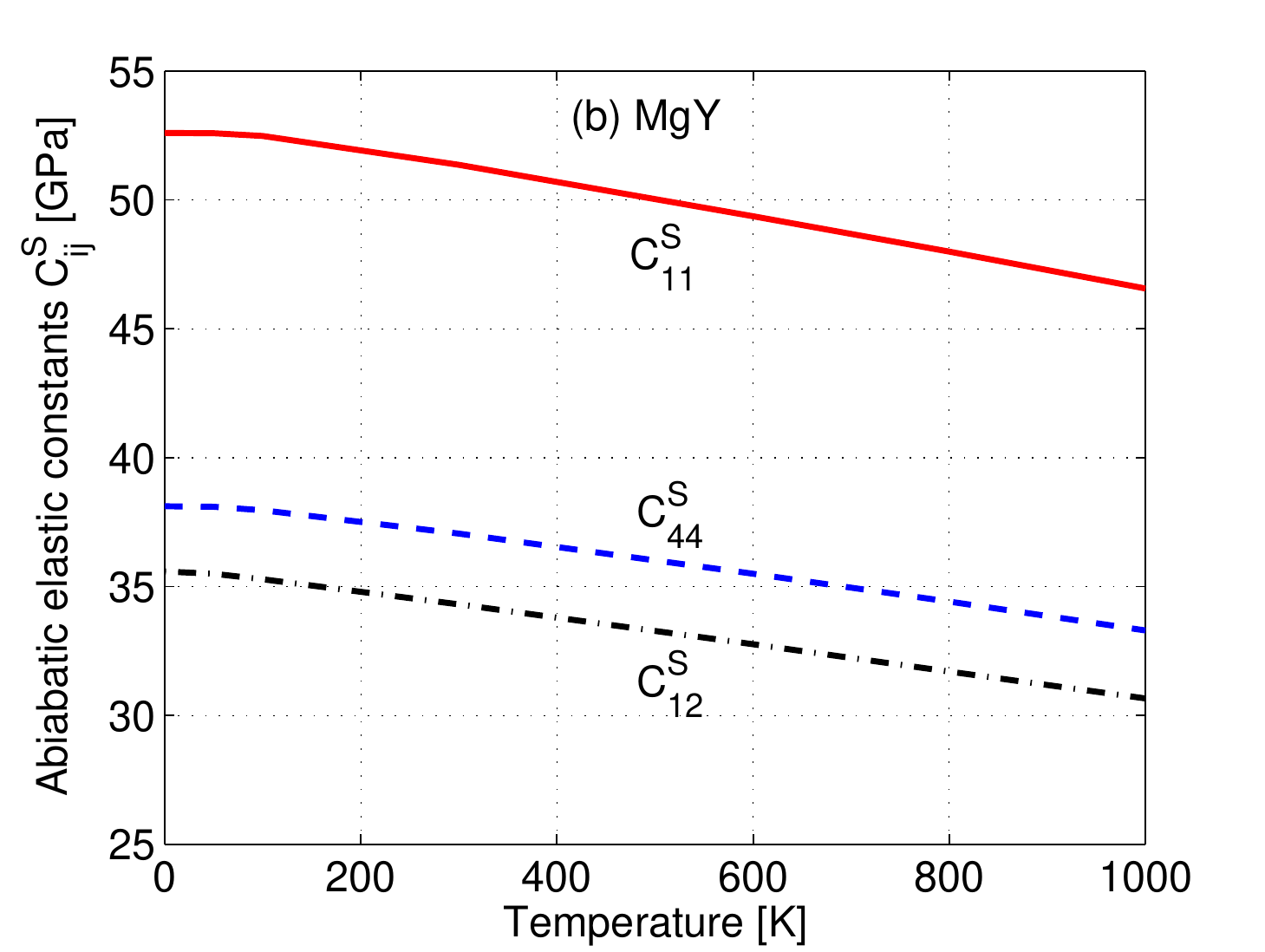}}
\scalebox{0.6}[0.6]{\includegraphics{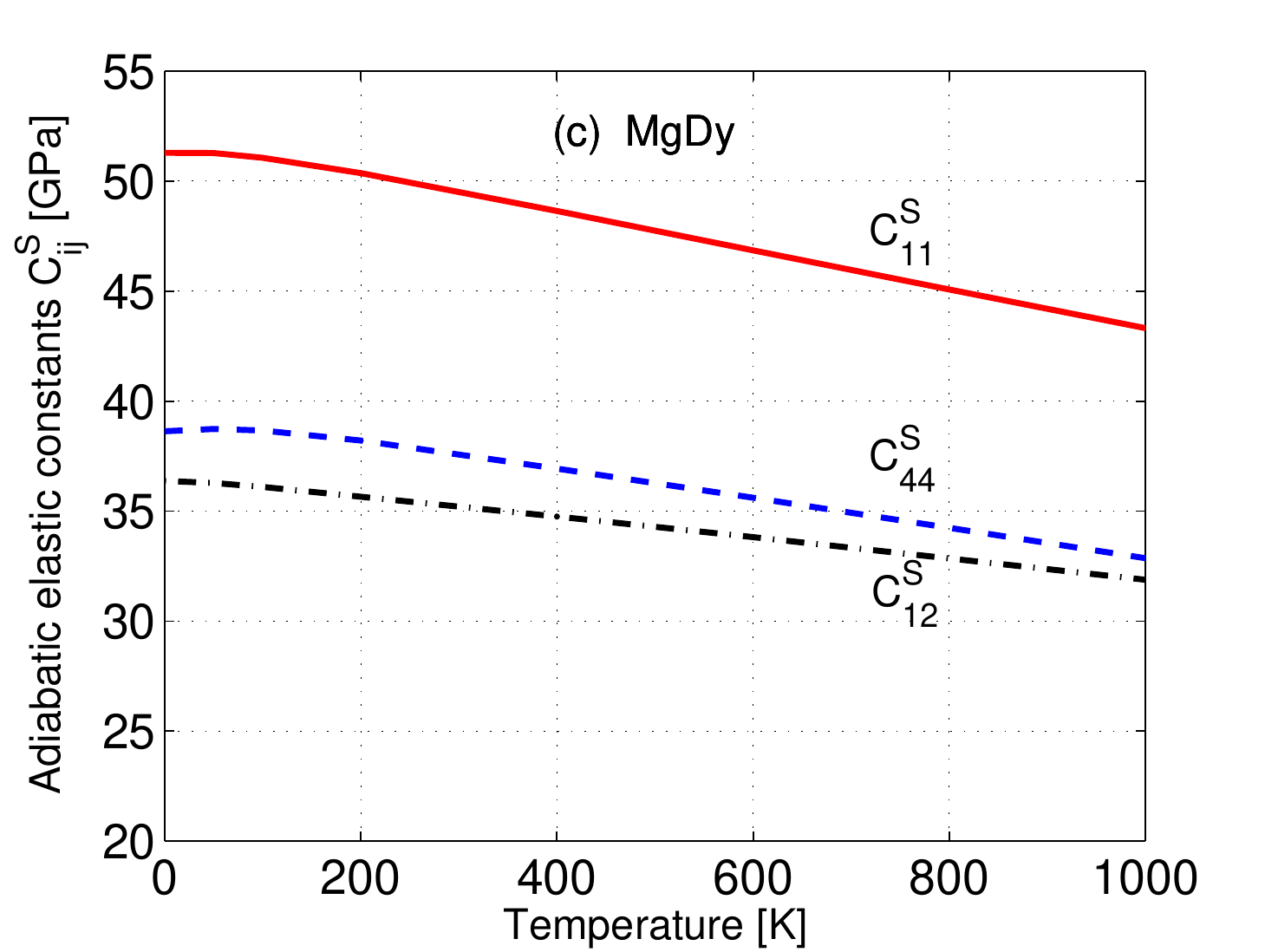}}
\scalebox{0.6}[0.6]{\includegraphics{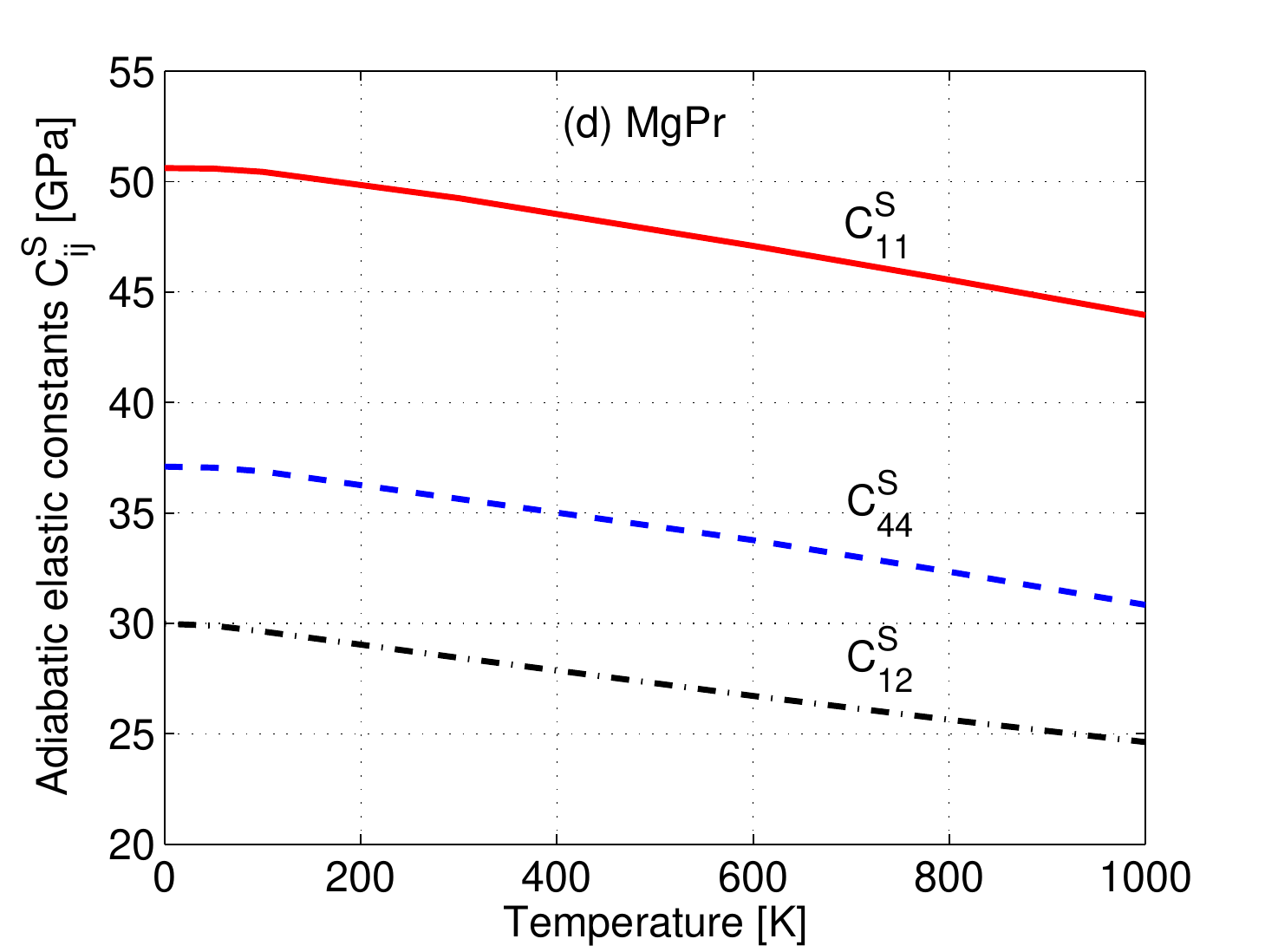}}
\scalebox{0.6}[0.6]{\includegraphics{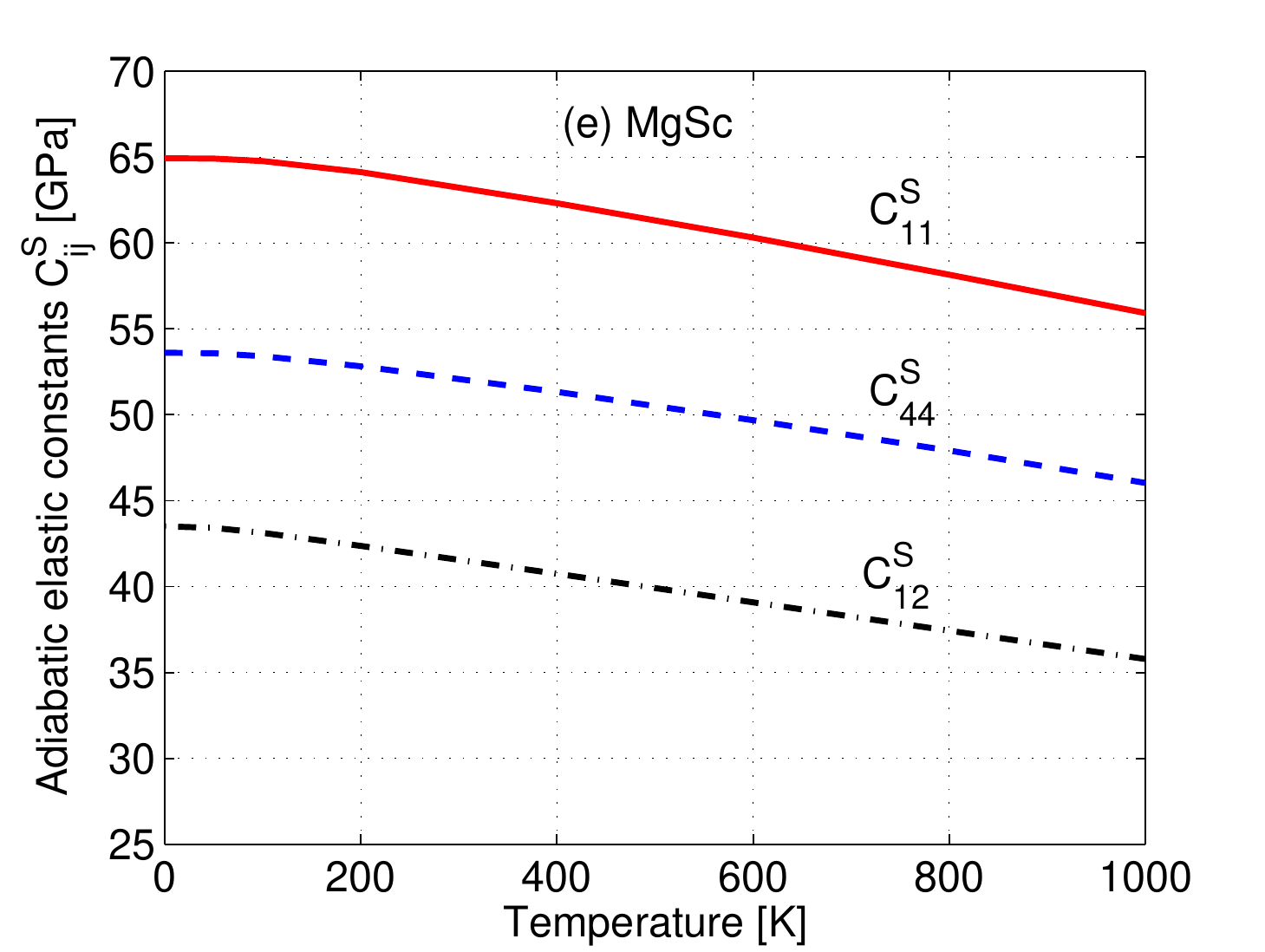}}
\scalebox{0.6}[0.6]{\includegraphics{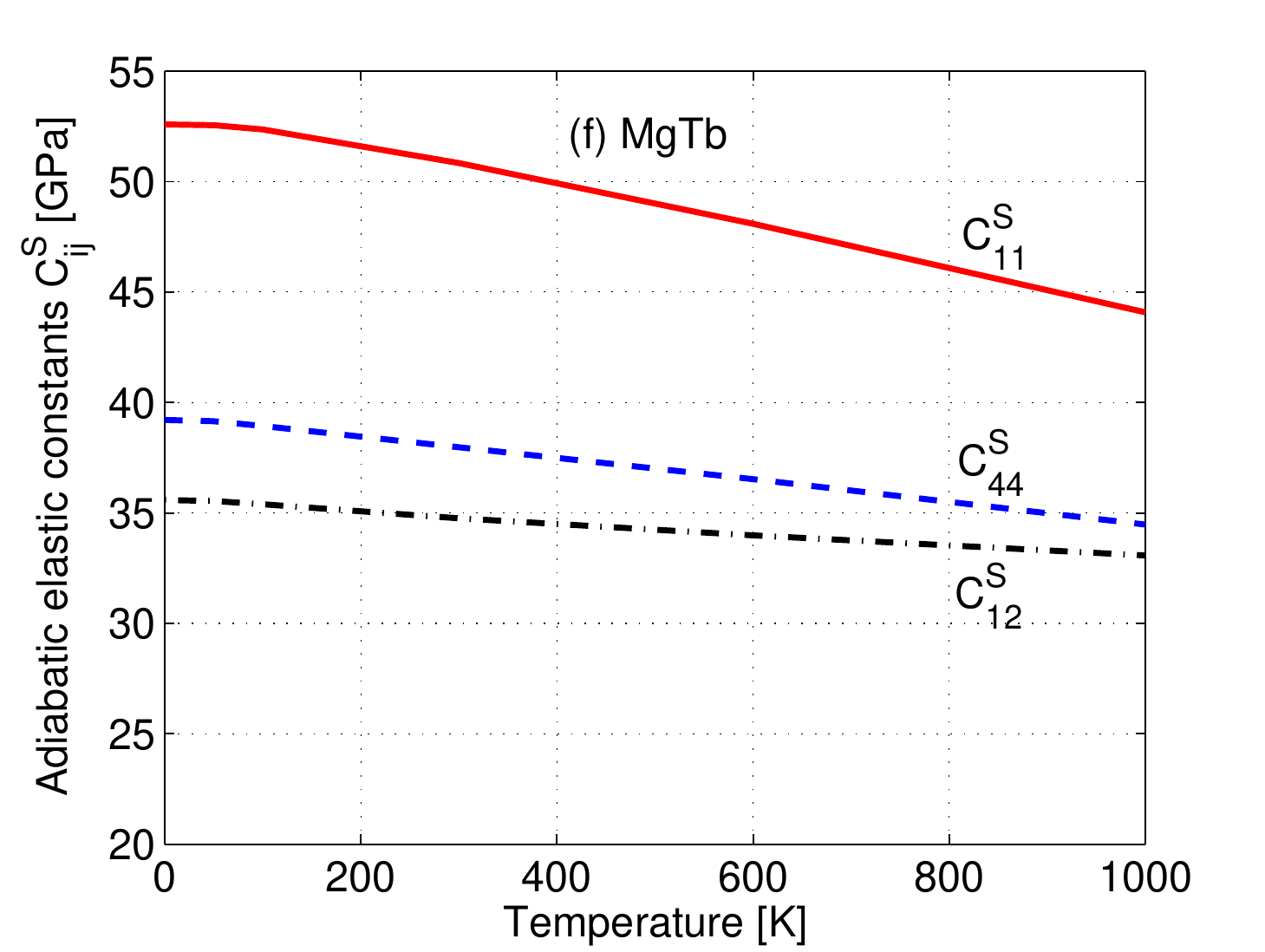}}
\caption{\footnotesize {(Color on line) The temperature-dependent adiabatic elastic constants for (a) NiAl, (b) MgY, (c) MgDy, (d) MgPr, (e) MgSc, and (f) MgTb. The solid, dashed-dotted, and dashed curves represents the  values of $C_{11}^{S}$, $C_{12}^{S}$ and $C_{44}^{S}$, respectively. For NiAl, the full symbols denote the corresponding experimental values form Rusovi\'{c} and Warlimont \cite{Rusovic}}. }
\label{elastic}
\end{figure}

Figure \ref{elastic} shows that the calculated temperature dependences of the adiabatic elastic constants $C_{11}^{S}$, $C_{12}^{S}$ and $C_{44}^{S}$ as a function of temperature. In Figure \ref{elastic} (a), we compare the calculated values of  $C_{ij}^{S}$ for NiAl with experiments \cite{Rusovic} and get a useful test of the accuracy of the method and the precision of our calculations. For all calculated MgRE intermetallics, we find the elastic constants $C_{ij}^{S}$ decrease with increasing temperature, since thermal expansion may soften the elastic moduli at high T. We also find that the trend of $C_{ij}^{S}$ approach linearity at higher temperature and zero slop around zero temperature. The calculated curves from Figure \ref{elastic} indicate that the elastic constants of MgRE follow a normal behavior with temperature. For MgY [Figure \ref{elastic}(b)], MgPr [Figure \ref{elastic}(d)], and MgSc [Figure \ref{elastic}(e)], we find that the curves of $C_{11}^{S}$, $C_{12}^{S}$ and $C_{44}^{S}$ are nearly  parallel, i.e., almost decrease to same extent in the whole temperature range. For MgDy [Figure \ref{elastic}(c)] and MgTb [Figure \ref{elastic}(f)], $C_{11}^{S}$ decrease to slightly larger than  $C_{12}^{S}$ and $C_{44}^{S}$. This character indicates that  MgRE intermetallics can keep their mechanical properties through wide temperature range and have good high temperature stability.

\section{Conclusions}

In summary, we have performed first-principles quasistatic approach to study the temperature-dependence of the elastic moduli of five MgRE (RE= Y, Dy, Pr, Sc, Tb) intermetallic compounds  with B2-type structure. Our calculations are based on the fact that the elastic moduli as a function of temperature  mainly results from thermal expansion. The elastic constants as function of volume are determined by the first-principles DFT total-energy calculations within the framework of the method of homogeneous deformation and the thermal expansion is determined by first-principles phonon calculations based on DFPT with QHA. To benchmark
the reliability results of the presented method, the comparison between the predicted results and the available experimental data for a benchmark material NiAl provides good agreements.  At $T=0K$, our calculated values of lattice parameter and elastic moduli for MgRE intermetallics show excellent agreement with previous theoretical results and experimental data. We find that the elastic constants follow a normal behavior with temperature: decrease with increasing temperature and approach linearity at higher temperature and zero slope around zero temperature.


 \vskip 2in

\footnotesize

\def\refname{{\large\bfseries References}}

\newpage

\begin{table}\tiny
\caption{\footnotesize {Our calculated lattice constants $a$, elastic constants $C_{ij}$, and bulk modulus $B$  for B2-MgRE (RE=Y, Dy, Pr, Sc, Tb) at $T=0K$ compared to previous computed results and experimental data. Note that $C_{ij}^{S}=C_{ij}^{T}=C_{ij}$ at $T=0K$.}}
\begin{tabular}{cccccc}
  \hline
   & MgY & MgDy & MgPr & MgSc & MgTb \\
   \hline
  $a$ ({\AA})    & 3.795$^{a}$, 3.796$^{b}$, 3.795$^{c}$, 3.796$^{d}$ &3.778$^{a}$, 3.765$^{b}$, 3.776$^{c}$, 3.759$^{d}$ & 3.910$^{a}$, 3.901$^{b}$, 3.909$^{c}$, 3.912$^{d}$ & 3.580$^{a}$, 3.593$^{b}$, 3.580$^{c}$, 3.597$^{d}$ & 3.789$^{a}$, 3.781$^{b}$, 3.789$^{c}$ 3.781$^{d}$ \\
  $C_{11}$ (GPa) & 52.60$^{a}$, 53.37$^{b}$, 53.07$^{c}$ & 51.29$^{a}$, 51.86$^{b}$, 51.85$^{c}$ & 50.61$^{a}$, 49.96$^{b}$, 49.45$^{c}$ & 64.93$^{a}$, 70.77$^{b}$, 63.86$^{c}$ & 52.59$^{a}$, 53.32$^{b}$, 52.46$^{c}$ \\
  $C_{12}$ (GPa) & 35.58$^{a}$, 36.39$^{b}$, 36.10$^{c}$ & 36.38$^{a}$, 37.57$^{b}$, 36.85$^{c}$ & 30.03$^{a}$, 30.80$^{b}$, 31.08$^{c}$ & 43.51$^{a}$, 43.11$^{b}$, 45.09$^{c}$ & 35.58$^{a}$, 36.18$^{b}$, 36.20$^{c}$ \\
  $C_{44}$ (GPa) & 38.12$^{a}$, 39.05$^{b}$, 39.26$^{c}$ & 38.63$^{a}$, 38.35$^{b}$, 39.74$^{c}$ & 37.10$^{a}$, 36.80$^{b}$, 36.20 & 53.61$^{a}$, 55.65$^{b}$, 52.62$^{c}$ & 39.21$^{a}$, 39.82$^{b}$, 39.95$^{c}$ \\
  $B$ (GPa)      & 41.25$^{a}$, 42.06$^{b}$, 41.75$^{c}$ & 41.35$^{a}$, 42.32$^{b}$, 41.85$^{c}$ & 36.86$^{a}$, 37.20$^{b}$, 37.20$^{c}$ & 50.65$^{a}$, 52.33$^{b}$, 51.35$^{c}$ & 41.25$^{a}$, 41.89$^{b}$, 41.62$^{c}$ \\
  \hline
\end{tabular}

\begin{tabular}{c}
  \leftline {${}^{a}$ This work;} \\
  \leftline {${}^{b}$ Ref. \cite{Wu} from first-principles calculations;}  \\
  \leftline {${}^{c}$ Ref. \cite{Tao} from first-principles calculations;} \\
  \leftline {${}^{d}$ Ref. \cite{Villars} from experiment.}\\
\end{tabular}
\label{table}
\end{table}

\begin{thebibliography}{5}
\bibitem{Kainer} K. U. Kainer, Magnessium Alloys and Their
Applications(Weinheim: Wiley-VCH)(2000).
\bibitem{Mordike1} B. L. Mordike, Mater. Sci. Eng. A 324( 2002) 103.
\bibitem{Mordike2} B. L. Mordike, J. Mater. Process. Technol. 117 (2001)
381.
\bibitem{Lorimer} G. W. Lorimer, P. J. Apps, H. Karimzaden, J. F. King,
Mater. Sci. Forum 419-422 (2003) 279.

\bibitem{Luca} S. E. Luca, M. Amara, R. M. Galera, J. F. Berar, J.
 Phys:Condens. Matter 14 (2002) 935.
\bibitem{Deldem} M.Deldem, M. Amara, R. M. Galera, P. Morin, D. Schmitt,
B. Ouladdiaf, J. Phys:Condens. Matter 10 (1998) 165.
\bibitem{Wang2010} R. Wang, S. F. Wang, X. Z. Wu, Y. Yao, and A. P. Liu, Intermetallics 18 (2010) 2472.
\bibitem{Wang2011p} X. Z. Wu, R. Wang, S. F. Wang, and L. L. Liu, Physica B 406 (2011) 967.
\bibitem{Wu} Y. Wu, W. Hu, Eur.Phys. J. B 60 (2007) 75.
\bibitem{Tao} X. M. Tao, Y. F. Ouyang, H. S. Liu, Y. P. Feng, Y. Du, Z.P. Jin, Solid State Communications 148 (2008) 314.
\bibitem{Wang2011} R. Wang, S. F. Wang, and X. Z. Wu, arXiv:1103.3146v1 (2011).
\bibitem{Guo} C. Guo, Z. Du, J. Alloys Compounds 422 (2006), p. 102.
\bibitem{Cacciamani} G. Cacciamani, S. De Negri, A. Saccone, R. Ferro, Intermetallics 11 (2003)
1135.
\bibitem{Wu1} Y. Wu, W. Hu, L. Sun, J. Phys. D: Appl. Phys. 40 (2007) 7584.
\bibitem{Gulsern} O. G\"{u}lsern and R. E. Cohen, Phys. Rev. B 65 (2002) 064103.
\bibitem{Karki} B. B. Karki, L. Stixrude, and R. M. Wentzcovitch, Rev. Geophys. 39 (2001) 507.
\bibitem{Orlikowski} D. Orlikowski, P. S\"{o}derlind, and J. A. Moriarty, Phys. Rev. B 74 (2006) 054109.
\bibitem{Kadas} K. K\'{a}das, L. Vitos, R. Ahuja, B. Johansson, and J. Koll\'{a}r, Phys. Rev. B 76 (2007) 235109.
\bibitem{Sha} X. W. Sha, and R.E. Cohen, Phys. Rev. B 81 (2010) 095105.
\bibitem{Wang} Y. Wang, J. J. Wang, H. Zhang, V. R. Manga, S. L. Shang, L. Q. Chen, and Z. K. Liu, J. Phys.: Condens. Matter. 22 (2010) 225404.
\bibitem{Shang} S. L. Shang, H. Zhang, Y. Wang, and  Z. K. Liu, J. Phys.: Condens. Matter 22 (2010) 375403.


\bibitem{Moriarty} J. A. Moriarty, J. F. Belak, R. E. Rudd, P. S\v{o}erlind, F. H. Streitz,
and L. H. Yang, J. Phys.: Condens. Matter 14 (2002) 2825 .
\bibitem{Wang1} Y. Wang, Z. K. Liu, L. Q. Chen, Acta Mater. 52 (2004) 2665.

\bibitem{Wasserman} E. Wasserman, L. Stixrude, and R.E. Cohen, Phys. Rev. B 53 (1996) 8296.
\bibitem{Brugger}K. Brugger. Phys Rev 133(1964) A1611.
\bibitem{Vinet} P. Vinet, J. H. Rose, J. Ferrante, J. R. Smith, J. Phys.: Condens. Matter 1 (1989) 1941.
\bibitem{Davies} G. F. Davies, J. Phys. Chem. Solids 35 (1974) 1513.



\bibitem{Kresse1} G. Kresse, J. Hafner, Phys. Rev. B 48 (1993) 3115.
\bibitem{Kresse2} G. Kresse, J. Furthm¨¹ller,  Comput. Mater. Sci. 6 (1996) 15.
\bibitem{Kresse3} G. Kresse, J. Furthm¨¹ller,  Phys. Rev. B 54 (1996) 11169.
\bibitem{Kresse5} G. Kresse, M. Marsman, J. Furthm\"{u}ller, VASP the guide, http://cms.mpi.univie.ac.at/vasp/.
\bibitem{Perdew1} J. P. Perdew, K. Burke, M. Ernzerhof, Phys. Rev. Lett. 77 (1996) 3865.
\bibitem{Perdew2} J. P. Perdew, K. Burke, M. Ernzerhof, Phys. Rev. Lett. 78 (1996) 1396.
\bibitem{Blochl} P. E. Bl$\ddot{o}$chl, Phys. Rev. B 50 (1994) 17953.
\bibitem{Kresse4} G. Kresse, D. Joubert, Phys. Rev. B 59 (1999) 1758.
\bibitem{Monkhorst} H. J. Monkhorst, J. D. Pack, Phys. Rev. B 13 (1976) 5188.
\bibitem{Togo2010} A. Togo, L. Chaput, I. Tanaka, G. Hug, Phys Rev B 81 (2010) 174301.
\bibitem{Togo2008}A. Togo, F. Oba, I. Tanaka, Phys. Rev. B, 78 (2008) 134106.
\bibitem{Togop} A. Togo, Phonopy, http://phonopy.sourceforge.net/.


\bibitem{Villars} P. Villars, L. D. Calvert, Pearson's Handbook of
Crystallographic Data for Intermetallic Phases (ASM, Metals Park,
oH, 1985).

\bibitem{Rusovic} N. Rusovi\'{c}, and H. Warlimont, Phys. Status Solid a 44 (1977) 609.
\bibitem{Nye} J. F. Nye, \textit{Physiccal Properties of Crystals} (Clarendon Press, Oxford, 1964).


\end{thebibliography}
\end{document}